\documentclass{egpubl}
\usepackage{envirvis2017-short}

\WsPaper         
\ifpdf \usepackage[pdftex]{graphicx} \pdfcompresslevel=9
\else \usepackage[dvips]{graphicx} \fi

\PrintedOrElectronic

\usepackage{t1enc,dfadobe}

\usepackage{egweblnk}
\usepackage{cite}
\usepackage{color}
\usepackage{titlesec}
\usepackage{verbatim}
\usepackage{enumitem}
\setlist[enumerate]{itemsep=0mm}

\titlespacing\subsubsection{0pt}{4pt plus 4pt minus 2pt}{0pt plus 2pt minus 2pt} 

\titlespacing\section{0pt}{8pt plus 4pt minus 2pt}{1pt plus 1pt minus 1pt}

\titlespacing\subsection{0pt}{4pt plus 4pt minus 2pt}{1pt plus 1pt minus 1pt}

\newcommand{\theName}{\textit{STOAViz}}

\title[\theName{}: Visualizing Saturated Thickness of Ogallala Aquifer]%
      {\theName{}: Visualizing Saturated Thickness of Ogallala Aquifer}

\author[T. Dang, L. Nguyen, A. Karim \& V. Uddameri]
      {
      T. Dang\thanks{Assistant professor, Computer Science department, Texas Tech University}
      L. Nguyen \thanks{PhD student, Computer Science department, Texas Tech University}
      A. Karim \thanks{PhD candidate, Civil Engineering department, Texas Tech University}
      V. Uddameri \thanks{Professor and director of Water Resources center, Texas Tech University}
      }

\begin{document}


\maketitle

\begin{abstract}
In this paper, we introduce \theName{}, a visual analytics tool for analyzing the saturated thickness of the Ogallala aquifer. The saturated thicknesses are monitored by sensors integrated on wells distributed on a vast geographic area. Our analytics application also captures the trends and patterns (such as average/standard deviation over time, sudden increase/decrease of saturated thicknesses) of water on an individual well and a group of wells based on their geographic locations. To highlight the usefulness and effectiveness of \theName{}, we demonstrate it on the Southern High Plains Aquifer of Texas. The work was developed using feedback from experts at the water resource center at a university. Moreover, our technique can be applied on any geographic areas where wells and their measurements are available.


\end{abstract}

\section{Introduction}

Water is the basic element that humans rely on for all living and manufacturing activities. According to the National Ground Water Association report in 2016 ~\cite{NGWA2016facts}, we extract around $982$ $km^3$ of water per year from the ground. This number serves as a good indicator of the need of water in daily life. Therefore, monitoring ground water, sustaining aquifer capability, and analyzing its changes are highly desirable by decision makers. Venki et al.~\cite{venki2017sensitivity} discussed two types of errors in well management: improper removal of wells and redundancy of insensitive wells. Moreover, as the number of wells increases, it becomes more and more challenging for specialists to visualize and analyze such a large amount of well monitoring data which comes in real time. 

In this paper, we propose a visual analytics tool for analyzing the saturated thickness of the Ogallala aquifer, called \theName{}. Saturated thickness is the vertical thickness of the hydro-geologically defined aquifer in which the pore spaces of the rock forming the aquifer are filled with water~\cite{Schloss2000current}.
The main contributions of this paper are:
\begin{itemize}[noitemsep,nolistsep]
\item We provide a data analytics tool for visualizing and analyzing a large number of well data distributed on a vast geographic area. Our approach is implemented as a web application which can track user location in real time.
\item We integrate time series features for detecting trends/patterns of monitored data at wells. This provides policy makers to come up with quick decisions.
\item We demonstrate our application in the Southern High Plains Aquifer of Texas and conduct an informal user study at a water resource center.
\end{itemize} 

The rest of this paper is organized as follows: We discuss the related work in the next section. Then we provide an overview of visualization tasks and describe implemented components in our system. We discuss our implementation and availability of our tool in Section~\ref{sec:4}. We provide the expert feedback from our user study and discussion in Section~\ref{sec:5}. Finally, we conclude our paper with future plans. 

\section{Related Work}
This section does not intend to survey all visualization tools for geospatial temporal visualizations~\cite{bach2014review}. Instead, we discuss some related tools. 
A simple approach to visualize temporal data on a 2D map is to attach the time series graphs directly on top of each geolocation\cite{andrienko2004interactive}. This easily becomes too clustered for larger time series, especially when geolocations are not equally distributed. Showing summary statistics~\cite{dang2013timeexplorer}, such as average, standard deviation, or trends~\cite{Shanbhag2005Temporal} on a choropleth map resolves this concern. However, this comes back to the trade-off between details and simplifications.  

\begin{figure*}[!ht]
 \centering
 \includegraphics[width=515pt]{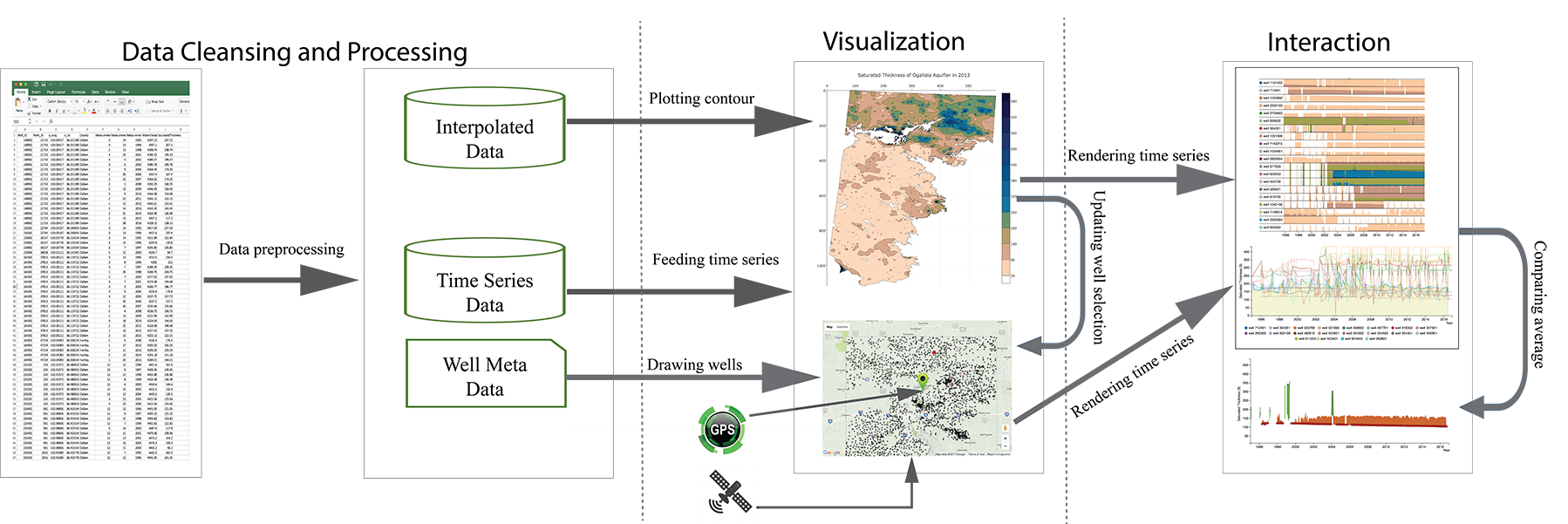}
 \caption{\label{fig:stages} Major stages in \theName{} visualization: data cleansing and processing, visualization, and interaction.}
\end{figure*}

Numerous recent visualization ideas have been proposed to address the challenges of big spatial and temporal data analytics. Keim et al.~\cite{Keim2004Infor-5420} suggest a three step process for information visualization: overview first, zoom and filter, and then details-on-demand. CrimeViz~\cite{roth2010user} utilizes this framework but also combines with mashup techniques to integrate additional visual elements on top of google map for crime analysis. Similarly, Ramakrishna et al.~\cite{ramakrishna2013interactive} present another software mashup which can handle a large spatio-temporal data set with 2.5 million records using hexagon binning. 

Keim et al.~\cite{keim2002information} propose   a classification of information visualization and visual data mining techniques which is based on the data type, the visualization technique, and the interaction technique. Soon after, Andrienko et al.~\cite{andrienko2003exploratory} provide an in-depth analytical review of exploratory spatio-temporal visualizations. The authors also categorize existing visualization-based techniques into two different types based on: (1) what types of spatio-temporal data they are applicable to; (2) what exploratory tasks they can potentially support.  Even though many techniques use spatial location, color, shape, and size to encode information, they do not generate useful insights by just using simple statistical summaries, such as average or standard deviation~\cite{andrienko2004interactive}. 

McCaffrey et al. ~\cite{mccaffrey2005unlocking} discuss various digital technologies for geological visualization and show a complete workflow from the initial data acquisition stage to the final project output. Wu et al. ~\cite{wu2004three} introduce feature-based representation to visualize three dimensional geological data. Hewagamage et al. ~\cite{hewagamage1999interactive} present spatial temporal visualization technique using spirals on geographical map. GeoTime~\cite{kapler2005geotime} enables users to track events, objects, and activities via a single interactive three dimensional view. Moving beyond traditional 2D displays, Mathiesen et al. ~\cite{mathiesen2012geological} applies Augmented Reality with generic smart phones and tablets to view existing geological data sets. 

Mohamed et al.~\cite{mohamed2013using} discuss the use of interactive visualizations for knowledge discovery in order to support decision making in medical domains. In the water resources domain, Venki et al.~\cite{venki2017sensitivity} combine ArgGIS 10.5~\cite{Env2016ArcGIS} with R integration~\cite{R2013R} for well data visualization and analysis. This requires strong technical skills from analysts and present limitations on interactive capabilities while analyzing multi-dimensional data.

\section{\theName{} Stages}
\theName{} aims to provide a high-level overview of saturated thickness of Ogallala aquifer. This tool also stretches out unusual patterns of water level such as sudden monthly increase/decrease~\cite{Hochheiser2004DQT,dang2013timeexplorer} and  overall trends~\cite{Buono2007Similarity}. This section explains \theName{} stages in detail. Figure~\ref{fig:stages} shows a schematic overview of \theName{}: 

\begin{enumerate}[noitemsep,nolistsep]
\item \textbf{Processing:} We first clean and preprocess the monthly data collected from well integrated sensors from 1995 to 2016. At the end of the Processing stage, we have time series data, spatial interpolated data, and well metadata. (see Section \ref{sec:3.1})

\item \textbf{Visualization:} There are multiple linked views in our \theName{} visualization: the saturated thickness contour map, the google map, the standard time series graph, and the horizon graphs~\cite{Reijner2008the} (see Section \ref{sec:3.2}).

\item \textbf{Interaction:} Users can click on a well or location on the map. The selected well (and other neighboring wells) are plotted in the time series view.  Users can further request a comparison chart to visualize individual well data and compare to its county average time series (see Section \ref{sec:3.3}).
\end{enumerate}   

The \theName{} implements five low-level analysis tasks that largely capture users' activities when working with a given dataset~\cite{Amar2005,Keim2004Infor-5420,andrienko2003exploratory}:
\begin{itemize}[noitemsep,nolistsep]
\item \textbf{T1}: Provide an overview of large number of well data.
\item \textbf{T2}: Retrieve and display well details on demand. 
\item \textbf{T3}: Group wells based on their geolocations.
\item \textbf{T4}: Filter and sort wells by their time series features~\cite{Hochheiser2004DQT}.
\item \textbf{T5}: Detect suspicious or abnormal wells.
\end{itemize} 

\subsection{Input datasets}
\label{sec:3.1}
We use the saturated thicknesses calculated at over 5,200 wells between the years 1995 and 2016 in the Southern High Plains Aquifer of Texas for visualization purposes. Saturated thickness in an unconfined aquifer is calculated using the equation: Saturated thickness (ST) = Land surface datum (LSD) $-$ Water level (WL) $-$ Bottom elevation of the aquifer (BT). LSD and BT are measured with respect to mean sea level (MSL), and WL is below ground surface.

\subsection{\theName{} overview}
\label{sec:3.2}

Main visualization components of \theName{} include the saturated thickness contour map, the google map, the standard time series chart, the horizon graphs, and the average comparison chart. All of these components share the same encodings: wells are drawn as filled circles whose radius is computed based on user selection strategy/feature. Selected wells are highlighted in different colors while other wells are in black with lower opacity.



\begin{figure}[!h]
 \centering
 \includegraphics[width=1\linewidth]{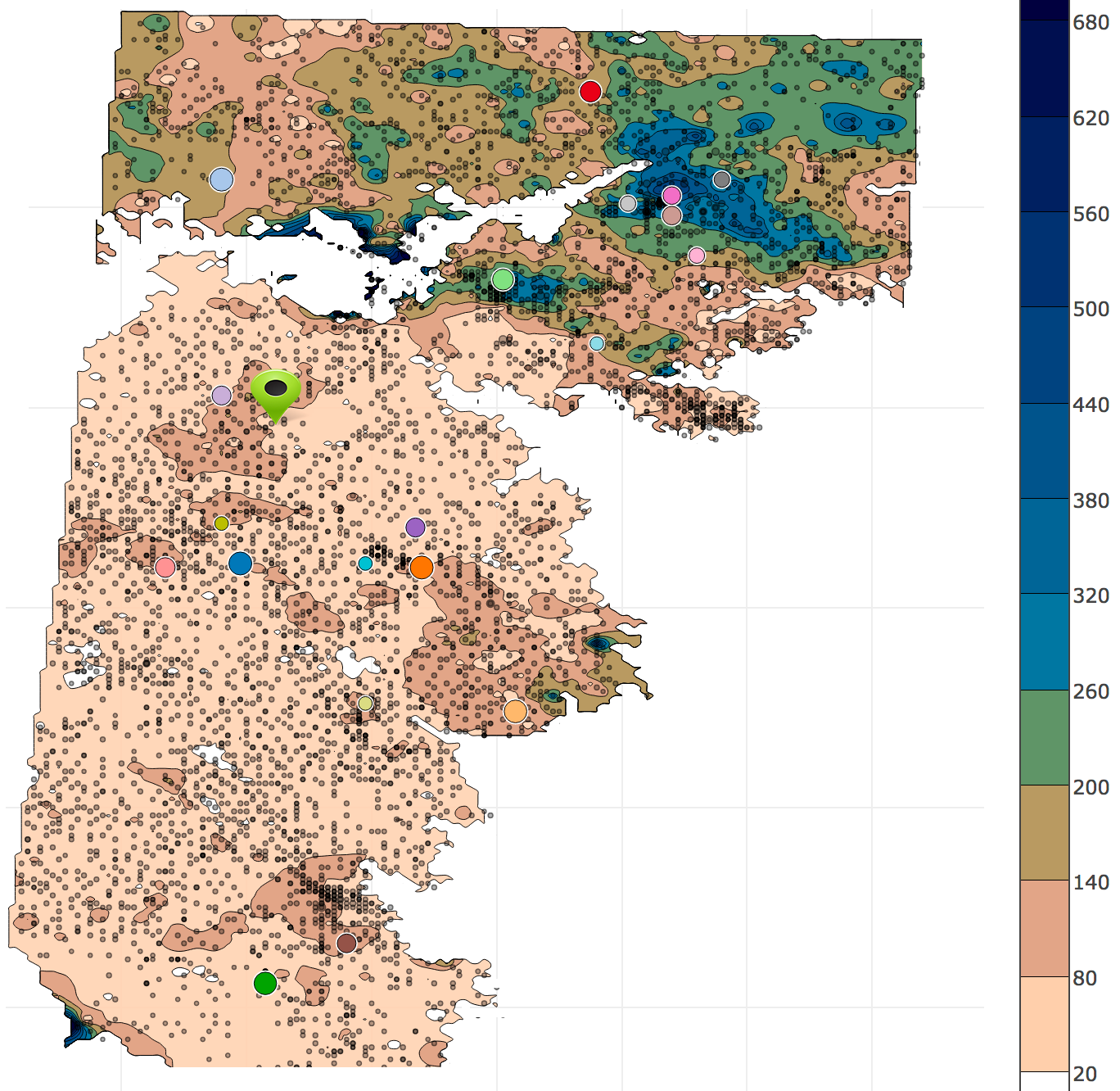}
 \caption{\label{fig:contour-map} Saturated thickness contour map of Ogallala in 2013. The green marker indicates user's current location.}
\end{figure}

\textbf{The Contour map} is plotted based on the well measurements and their interpolated data for areas where these measurements are not available. This map provides an high-level overview of water supply capability for a given geographic area (visualization task \textbf{T1}) as depicted in Figure~\ref{fig:contour-map}. In particular, the darker blue regions are areas with a high volume of water supply while the lighter brown regions have thinner saturated thicknesses. Wells are overlaid on top of this contour map. Colored circles are wells with ``highest monthly drop'' on their time series. 

Moreover, if a user has location tracking enabled via browser or mobile phone, a flickering filled circle is displayed to represent the current position of the user. This will help the user to figure out how much water capability exists in that area.

\begin{figure}[!h]
 \centering
 \includegraphics[width=1\linewidth]{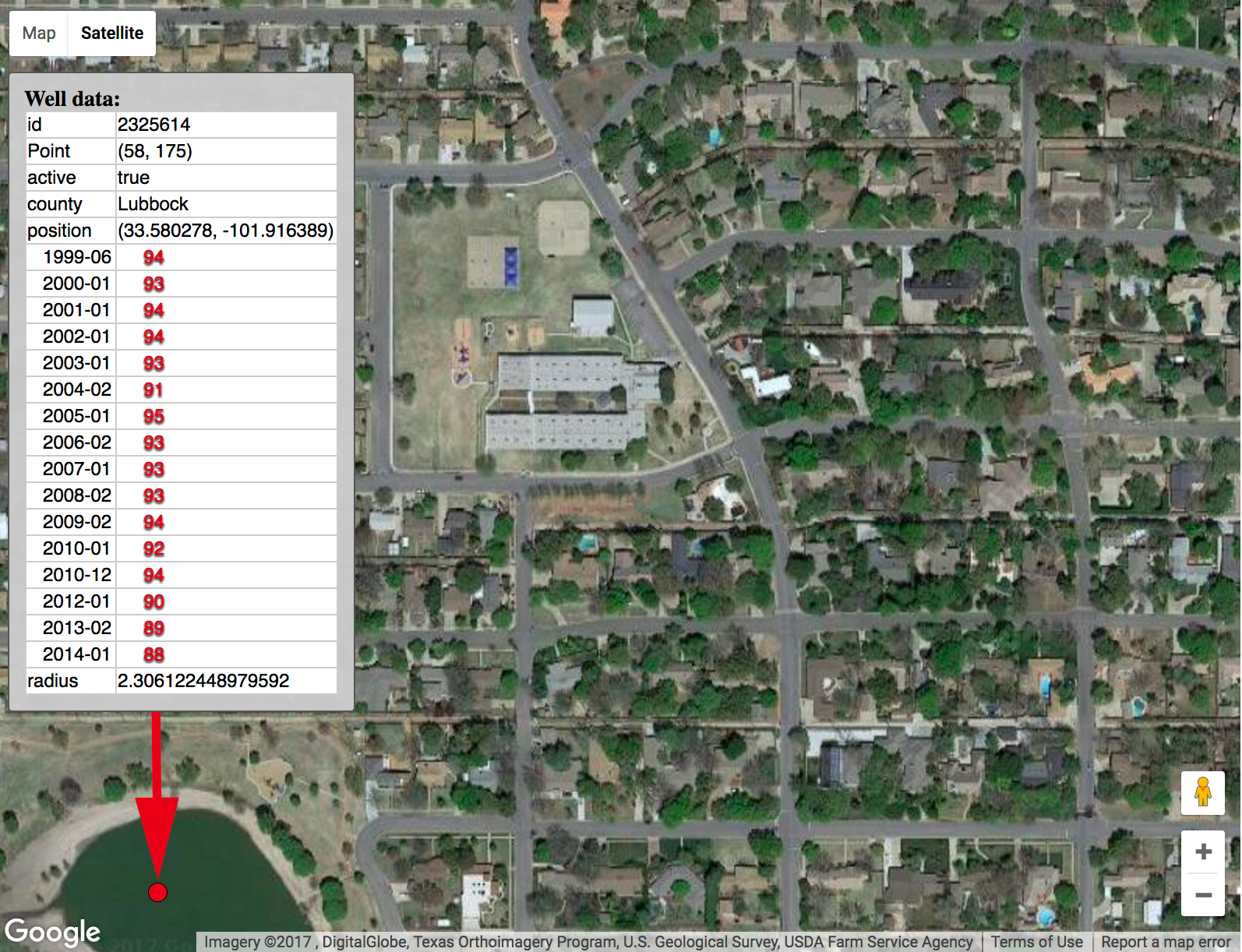}
 \caption{\label{fig:Figure3} Visualizing wells on google map (satellite view). The popup window displays well details on mouse over.}
\end{figure}

\textbf{The google map} The contour map serves as a starting point. As users select a well/location on the contour map, all of the neighboring wells (or all wells in the selected county) are plotted (visualization task \textbf{T3}) in the google map for further inspections. All google-map navigation capabilities, such as zooming in/out and panning, are also supported within this view. Figure~\ref{fig:Figure3} shows an example when a user mouses over a well (at the end of red arrow).  All information (well ID, GPS location, county, and measured saturated thickness over time) are displayed in a popup window (visualization task \textbf{T2}). This view is also linked to time series visualizations (the standard line graph or horizon graphs, based on user choice).

\begin{figure}[!h]
 \centering
 \includegraphics[width=1\linewidth]{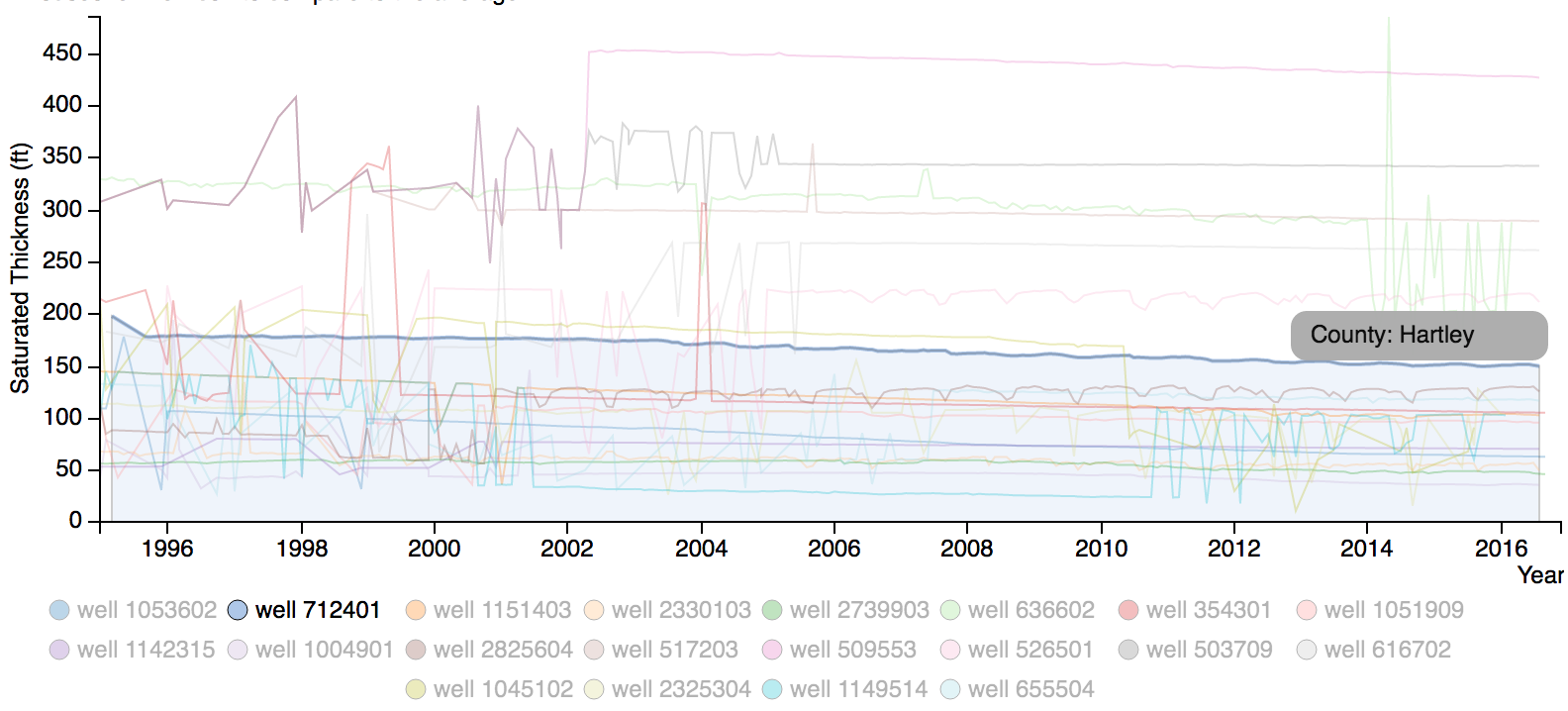}
 \caption{\label{fig:time-series} The standard line graphs for wells with a selected well in Hartley county. This selected well is further examined in Figure~\ref{fig:average-comparison}.}
\end{figure}

\textbf{The time series line graphs} for selected wells is depicted in Figure~\ref{fig:time-series}. We use the same color encoding across different views in \theName{}. When a user mouses over any well at the bottom legend, the selected well's line chart is highlighted while others are faded out. 

\begin{figure}[!h]
 \centering
 \includegraphics[width=1\linewidth]{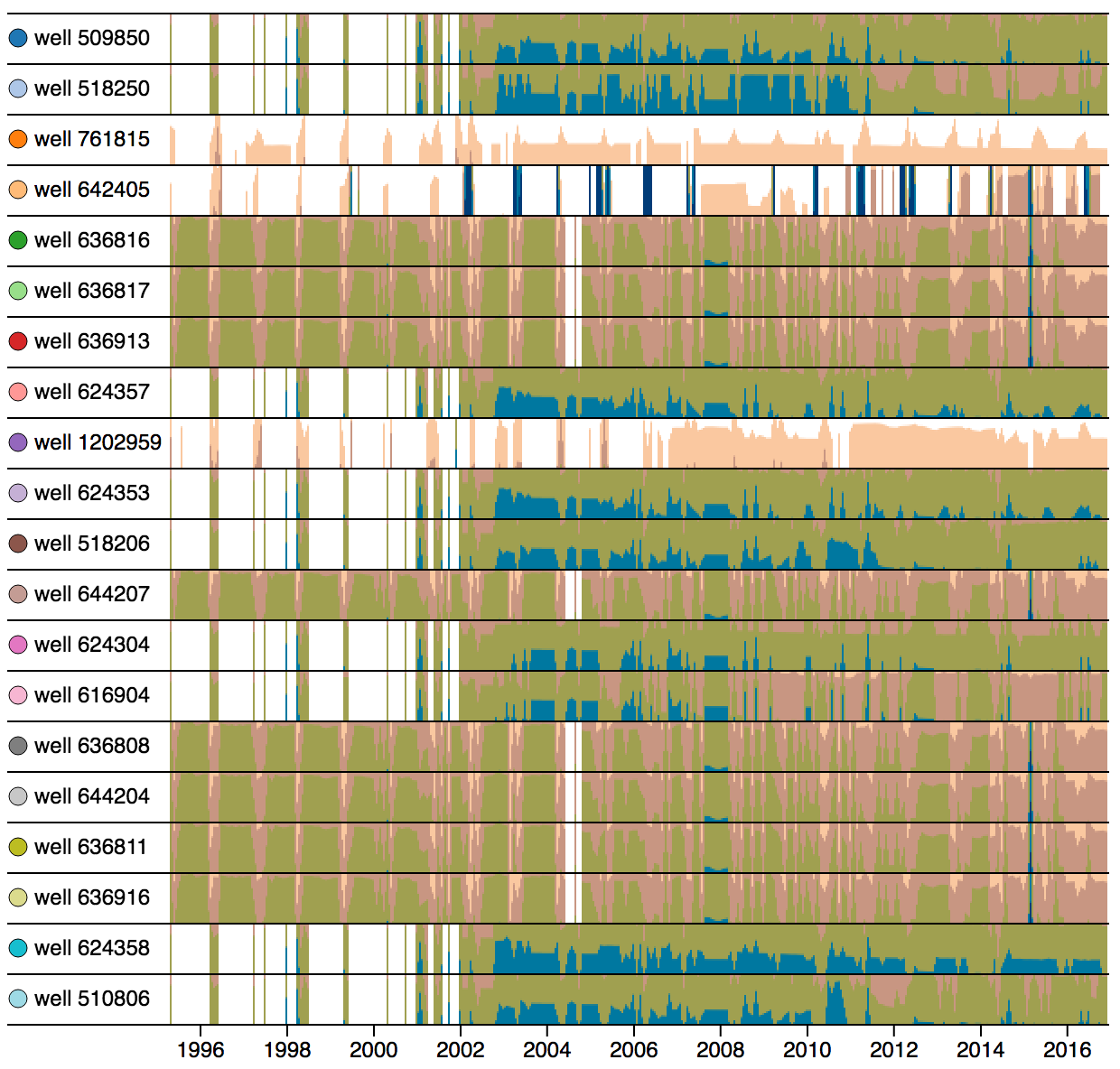}
 \caption{\label{fig:horizon} The horizon graphs of the same 20 wells as in Figure~\ref{fig:time-series}. Darker blue are thicker saturated thicknesses. }
\end{figure}

\textbf{The horizon graphs}
Users can switch to horizon graphs (instead of standard line graphs) as depicted in Figure~\ref{fig:horizon}. This graph provides a different perspective in viewing well time series~\cite{Javed2010BradedGraph}. In particular, the time series is split into bands using a uniform interval. The interval and band colors are the same as on the contour map (see the right color legend of Figure~\ref{fig:contour-map}). The split bands are now collapsed so that the higher bands layer on top~\cite{Reijner2008the}. The filled well time series chart becomes a horizon graph that takes up much less space~\cite{NathanHorizon}. (How much less depends on how many split bands that we want.)

The blank time intervals on the horizon graph are time periods when well measurements are not available. Blue bands represent high volumes of water supply. The wells in Figure~\ref{fig:horizon} have been ordered based on how much water they lost within two consecutive months (visualization task \textbf{T4}). Upon mousing over any wells, a comparison graph of this well to the average time series is displayed. The comparison graph is described in the next section.

\subsection{Interaction}
\label{sec:3.3}

\begin{figure}[!h]
 \centering
 \includegraphics[width=1\linewidth]{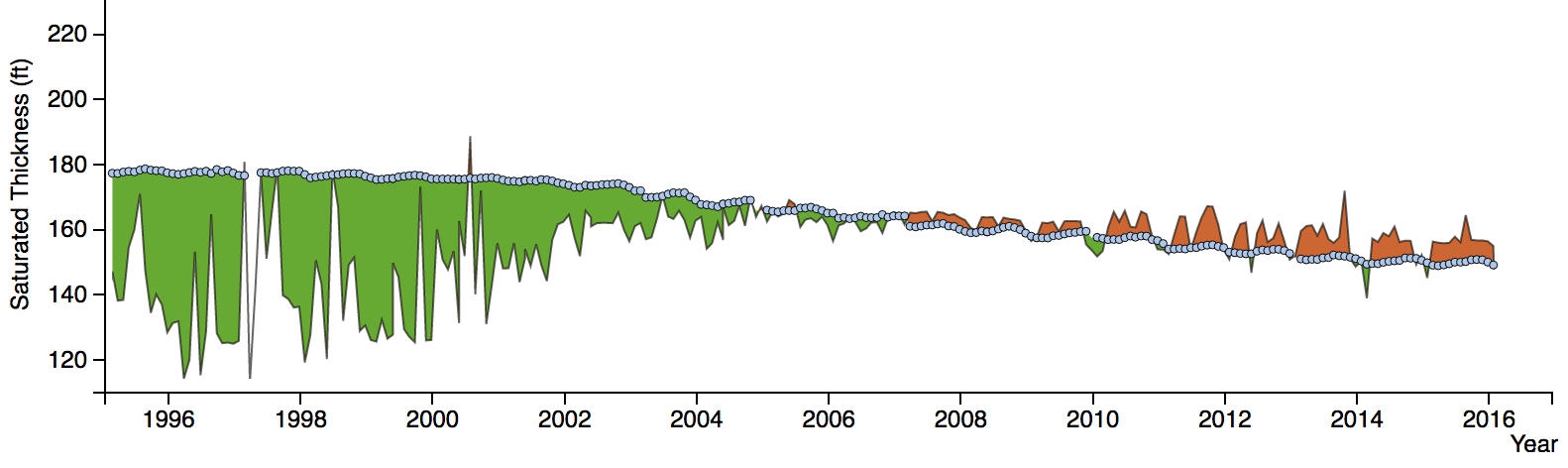}
 \caption{\label{fig:average-comparison} Well 712401 in Hartley county (light blue) and county average comparison (black line): green for above and orange for below the county average.}
\end{figure}

All views in \theName{} are linked. In other words, as users make a different selection on one view, others are updated accordingly. Moreover, users can request to show the comparison graph of a selected well to county/neighbor average time series. A bipolar color band is used to highlight differences: green for above and orange for below the county average. As depicted in Figure~\ref{fig:average-comparison}, the saturated thickness of well 712401 in Hartley county is becoming thinner and thinner (after 2008, the saturated thickness is constantly below the county average). Policy makers in Hartley county may want to take actions to stop this decreasing trend.

\section{Implementation}
\label{sec:4}

\theName{} is implemented as a web application utilizing the D3.js library~\cite{Bostock2011D3} and the Google map API ~\cite{Alphabet2008the}. This is an open source project. Source code, video, web demo, and project documentation are available on our Github repository at \url{https://github.com/iDataVisualizationLab/SaturatedThickness}.

\section{Evaluation and discussion}
\label{sec:5}
We solicited qualitative responses about \theName{} from two experts at a water resource center: one post-doc researcher and one associate professor. Both have at least nine years of experience in this research domain. The informal study started with a quick description (around 10 minutes) of the main components and their functionalities to familiarize users to the main GUI of \theName{}. Then the experts are free to use \theName{} before providing feedback. Both of them agree that the contour map provides a high level overview and can serve as the starting point for their well selection while the google map provides details of wells. (They can turn on the satellite images to inspect well location and county, see Figure ~\ref{fig:Figure3}.) Moreover, the average graph is useful to see how the saturated thickness at a particular well behaves compared to others in the same county. 

For the time series visualizations, they indicated that they are familiar with the line graphs but not the horizon graph. However, they both got the main idea of horizon graphs after a short explanation. Each expert has different interests and selected different features to inspect the time series data of wells. One selected wells with high variance of saturated thickness over time (depicted in Figure~\ref{fig:Figure3}). She indicated that with horizon graphs, she can quickly compare patterns of a large number of wells. It makes sense that neighboring wells on the map have similar patterns. Moreover, missing data (well measurements are not available for certain period of time) can be discerned quickly. 

The other expert selected ``highest monthly drop'' and detected an suspicious drop of water within a month (visualization task \textbf{T5}). He then used the Google map and grasp all wells within that county and many of them have similar drop on that month. He also commented that ``for a smaller selection of wells (fewer than 5), I would like to use standard line graph since I can read the monthly values easily. For a larger number of wells, line graph becomes too cluttered. Therefore, horizon patterns are much easier to compare, especially for wells with higher saturated thicknesses (blue areas).''  

Besides the positive feedback, the experts also pointed out some limitations of \theName{}. For example, they suggested that the Google map and contour map should be automatically updated via interactions with the time series visualizations. These features have been incorporated in the final product on our Github repository.

\section{Conclusion and Future work}
In this paper, we present \theName{}, a visual analytics tool with a case study of the Southern High Plains Aquifer of Texas. The tool utilizes state of art spatial-temporal visualization techniques to support various linked views of saturated thicknesses monitored by a large number of wells distributed on a vast geographic area. 

For future work, we will extend the application with Nanocubes integration ~\cite{lins2013nanocubes}. Nanocubes provides real-time visualization of billions of multidimensional spatial-temporal data expand zooming and other map viewing experiences.  While the current version of \theName{} is limited to one layer of underground water (the saturated thickness contour map in Figure~\ref{fig:contour-map}), we eventually target a 3D \theName{} which allows users to visualize and interact with different dynamic layers of underground water in a virtual reality environment. 

\bibliographystyle{eg-alpha}
\bibliography{bibliography}

\newcommand{\etalchar}[1]{$^{#1}$}
\begin{thebibliography}{\uppercase{MMAT12}}

\bibitem[AA04]{andrienko2004interactive}
\textsc{Andrienko N., Andrienko G.}:
\newblock Interactive visual tools to explore spatio-temporal variation.
\newblock In \emph{Proceedings of the working conference on Advanced visual
  interfaces} (2004), ACM, pp.~417--420.

\bibitem[AAG03]{andrienko2003exploratory}
\textsc{Andrienko N., Andrienko G., Gatalsky P.}:
\newblock Exploratory spatio-temporal visualization: an analytical review.
\newblock \emph{Journal of Visual Languages \& Computing 14}, 6 (2003),
  503--541.

\bibitem[AES05]{Amar2005}
\textsc{Amar R., Eagan J., Stasko J.}:
\newblock Low-level components of analytic activity in information
  visualization.
\newblock In \emph{Proc. of the IEEE Symposium on Information Visualization}
  (2005), pp.~15--24.

\bibitem[BDA{\etalchar{*}}14]{bach2014review}
\textsc{Bach B., Dragicevic P., Archambault D., Hurter C., Sheelagh C.}:
\newblock A review of temporal data visualizations based on space-time cube
  operations.
\newblock In \emph{Eurographics Conference on Visualization} (2014),
  Eurographics.

\bibitem[BOH11]{Bostock2011D3}
\textsc{Bostock M., Ogievetsky V., Heer J.}:
\newblock D3 data-driven documents.
\newblock \emph{{IEEE} Trans. Vis. Comput. Graph. 17}, 12 (2011), 2301--2309.

\bibitem[BPS{\etalchar{*}}07]{Buono2007Similarity}
\textsc{Buono P., Plaisant C., Simeone A., Aris A., Shmueli G., Jank W.}:
\newblock Similarity-based forecasting with simultaneous previews: A river plot
  interface for time series forecasting.
\newblock In \emph{Information Visualization, 2007. IV '07. 11th International
  Conference} (July 2007), pp.~191--196.

\bibitem[DW13]{dang2013timeexplorer}
\textsc{Dang T., Wilkinson L.}:
\newblock {TimeExplorer: S}imilarity search time series by their signatures.
\newblock In \emph{Proc. International Symp. on Visual Computing} (2013),
  pp.~280--289.

\bibitem[{Env}16]{Env2016ArcGIS}
\textsc{{Environmental Systems Research Institute}}:
\newblock \emph{ArcGIS Desktop: Release 10.5}.
\newblock ESRI, Redlands, California, CA, 2016.

\bibitem[HHI99]{hewagamage1999interactive}
\textsc{Hewagamage K.~P., Hirakawa M., Ichikawa T.}:
\newblock Interactive visualization of spatiotemporal patterns using spirals on
  a geographical map.
\newblock In \emph{Visual languages, 1999. Proceedings. 1999 IEEE symposium on}
  (1999), IEEE, pp.~296--303.

\bibitem[HR08]{Reijner2008the}
\textsc{Hannes~Reijner R. W.~B.}:
\newblock The development of the horizon graph.
\newblock
  \href{http://www.stonesc.com/Vis08\_Workshop/DVD/Reijner\_submission.pdf}{http://www.stonesc.com/Vis08\_Workshop/DVD/Reijner\_submission.pdf},
  2008.

\bibitem[HS04]{Hochheiser2004DQT}
\textsc{Hochheiser H., Shneiderman B.}:
\newblock Dynamic query tools for time series data sets: Timebox widgets for
  interactive exploration.
\newblock \emph{Information Visualization 3}, 1 (Mar. 2004), 1--18.

\bibitem[Inc]{Alphabet2008the}
\textsc{Inc. A.}:
\newblock Google map.
\newblock \href{http://maps.google.com}{http://maps.google.com}.

\bibitem[JAS00]{Schloss2000current}
\textsc{J.~A.~Schloss R. W.~B.}:
\newblock Current saturated thickness.
\newblock
  \href{http://www.kgs.ku.edu/HighPlains/atlas/atcst.htm}{http://www.kgs.ku.edu/HighPlains/atlas/atcst.htm},
  last revision 2000.

\bibitem[JME10]{Javed2010BradedGraph}
\textsc{Javed W., McDonnel B., Elmqvist N.}:
\newblock Graphical perception of multiple time series.
\newblock \emph{{IEEE} Trans. Vis. Comput. Graph. 16}, 6 (Nov. 2010), 927--934.

\bibitem[Kei02]{keim2002information}
\textsc{Keim D.~A.}:
\newblock Information visualization and visual data mining.
\newblock \emph{IEEE transactions on Visualization and Computer Graphics 8}, 1
  (2002), 1--8.

\bibitem[KPS04]{Keim2004Infor-5420}
\textsc{Keim D.~A., Panse C., Sips M.}:
\newblock Information visualization : Scope, techniques and opportunities for
  geovisualization.
\newblock In \emph{Exploring Geovisualization}, Dykes J., (Ed.). Elsevier,
  Oxford, 2004, pp.~1--17.

\bibitem[KW05]{kapler2005geotime}
\textsc{Kapler T., Wright W.}:
\newblock Geotime information visualization.
\newblock \emph{Information Visualization 4}, 2 (2005), 136--146.

\bibitem[LKS13]{lins2013nanocubes}
\textsc{Lins L., Klosowski J.~T., Scheidegger C.}:
\newblock Nanocubes for real-time exploration of spatiotemporal datasets.
\newblock \emph{IEEE Transactions on Visualization and Computer Graphics 19},
  12 (2013), 2456--2465.

\bibitem[MJH{\etalchar{*}}05]{mccaffrey2005unlocking}
\textsc{McCaffrey K., Jones R., Holdsworth R., Wilson R., Clegg P., Imber J.,
  Holliman N., Trinks I.}:
\newblock Unlocking the spatial dimension: digital technologies and the future
  of geoscience fieldwork.
\newblock \emph{Journal of the Geological Society 162}, 6 (2005), 927--938.

\bibitem[MLA13]{mohamed2013using}
\textsc{Mohamed E.~B., Ltifi H., Ayed M.~B.}:
\newblock Using visualization techniques in knowledge discovery process for
  decision making.
\newblock In \emph{Hybrid Intelligent Systems (HIS), 2013 13th International
  Conference on} (2013), IEEE, pp.~93--98.

\bibitem[MMAT12]{mathiesen2012geological}
\textsc{Mathiesen D., Myers T., Atkinson I., Trevathan J.}:
\newblock Geological visualisation with augmented reality.
\newblock In \emph{Network-Based Information Systems (NBiS), 2012 15th
  International Conference on} (2012), IEEE, pp.~172--179.

\bibitem[NGW16]{NGWA2016facts}
\textsc{NGWA}:
\newblock Facts about global groundwater usage.
\newblock Westerville, Ohio 43081-8978 USA.

\bibitem[{R C}13]{R2013R}
\textsc{{R Core Team}}:
\newblock \emph{R: A Language and Environment for Statistical Computing}.
\newblock R Foundation for Statistical Computing, Vienna, Austria, 2013.
\newblock {ISBN} 3-900051-07-0.

\bibitem[RCM13]{ramakrishna2013interactive}
\textsc{Ramakrishna A., Chang Y.-H., Maheswaran R.}:
\newblock An interactive web based spatio-temporal visualization system.
\newblock In \emph{International Symposium on Visual Computing} (2013),
  Springer, pp.~673--680.

\bibitem[RRF{\etalchar{*}}10]{roth2010user}
\textsc{Roth R.~E., Ross K.~S., Finch B.~G., Luo W., MacEachren A.~M.}:
\newblock A user-centered approach for designing and developing spatiotemporal
  crime analysis tools.
\newblock In \emph{Proceedings of GIScience} (2010), vol.~15.

\bibitem[SRd05]{Shanbhag2005Temporal}
\textsc{Shanbhag P., Rheingans P., desJardins M.}:
\newblock Temporal visualization of planning polygons for efficient
  partitioning of geo-spatial data.
\newblock In \emph{IEEE Symposium on Information Visualization, 2005. INFOVIS
  2005.} (Oct 2005), pp.~211--218.

\bibitem[VUS17]{venki2017sensitivity}
\textsc{Venki~Uddameri Abdullah~Karim E.~U., Srivastava P.}:
\newblock Sensitivity of wells in a large groundwater monitoring network and
  its evaluation using grace satellite derived information.
\newblock In \emph{Sensitivity Analysis in Earth Observation} (2017), Elsevier,
  pp.~235--256.

\bibitem[WX04]{wu2004three}
\textsc{Wu Q., Xu H.}:
\newblock On three-dimensional geological modeling and visualization.
\newblock \emph{Science in China Series D: Earth Sciences 47}, 8 (2004),
  739--748.

\bibitem[YAU]{NathanHorizon}
\textsc{YAU N.}:
\newblock Horizon graphs, with a food pricing example.
\newblock
  \href{https://flowingdata.com/2015/07/02/changing-price-of-food-items-and-horizon-graphs/}{https://flowingdata.com/2015/07/02/changing-price-of-food-items-and-horizon-graphs/}.

\end{thebibliography}

\end{document}